\documentclass[aps,prd,twocolumn,showpacs,groupedaddress,floatfix]{revtex4}

\usepackage{graphicx}
\DeclareGraphicsExtensions{.eps, .eps, .jpg, .png}

\def\by{{\bar{y}}}
\def\ba{{a\left(\by\right)}}

\def\be{\begin{equation}}
\def\ee{\end{equation}}
\def\ba{\begin{eqnarray}}
\def\ea{\end{eqnarray}}

\def\bdm{\begin{displaymath}}
\def\edm{\end{displaymath}}
\def\la{~\mbox{\raisebox{-.6ex}{$\stackrel{<}{\sim}$}}~}
\def\ga{~\mbox{\raisebox{-.6ex}{$\stackrel{>}{\sim}$}}~}
\def\bq{\begin{quote}}
\def\eq{\end{quote}}

 at 10truept

\newcommand{\beq}{\begin{equation}}
\newcommand{\eeq}{\end{equation}}
\newcommand{\bea}{\begin{eqnarray}}
\newcommand{\eea}{\end{eqnarray}}
\newcommand{\beqa}{\begin{eqnarray}}
\newcommand{\eeqa}{\end{eqnarray}}

\def\la{~\mbox{\raisebox{-.6ex}{$\stackrel{<}{\sim}$}}~}
\def\ga{~\mbox{\raisebox{-.6ex}{$\stackrel{>}{\sim}$}}~}

\def\ltap{\ \raise.3ex\hbox{$<$\kern-.75em\lower1ex\hbox{$\sim$}}\ }
\def\gtap{\ \raise.3ex\hbox{$>$\kern-.75em\lower1ex\hbox{$\sim$}}\ }
\def\gl{\ \raise.5ex\hbox{$>$}\kern-.8em\lower.5ex\hbox{$<$}\ }
\def\roughly#1{\raise.3ex\hbox{$#1$\kern-.75em\lower1ex\hbox{$\sim$}}}

\begin{document}



\title{Where in the String Landscape is Quintessence}


\author{Nemanja Kaloper}
\email[]{kaloper@physics.ucdavis.edu}
\affiliation{Department of Physics, University of
California, Davis, CA 95616, USA}

\author{Lorenzo Sorbo}
\email[]{sorbo@physics.umass.edu}
\affiliation{Department of Physics,
University of Massachusetts, Amherst, MA 01003, USA}


\date{\today}

\begin{abstract}

We argue that quintessence may reside in certain corners of the string landscape. 
It arises as a linear combination of internal 
space components of higher rank forms, which are axion-like at low energies, and
may mix with $4$-forms after compactification of the Chern-Simons terms to $4D$ due 
to internal space fluxes. The mixing 
induces an effective mass term, with an action which {\it preserves} the axion shift symmetry,
breaking it spontaneously after the background selection. With several axions, 
several $4$-forms, and a low string scale, as in one of the setups
already invoked for dynamically explaining a tiny residual vacuum 
energy in string theory, the $4D$ mass matrix generated by random fluxes may have ultralight 
eigenmodes over the landscape, which are quintessence. 
We illustrate how this works in simplest cases, 
and outline how to get the lightest mass to be 
comparable to the Hubble scale now, $H_0 \sim 10^{-33} {\rm eV}$.
The shift symmetry protects the smallest mass from perturbative corrections in field 
theory. Further, if the ultralight eigenmode does not couple directly to any sector strongly 
coupled at a high scale, the non-perturbative field theory corrections to its potential will also 
be suppressed. Finally, if the compactification length is larger than
the string length by more than an order of magnitude, the gravitational
corrections may remain small too, even when the field value approaches $M_{Pl}$.

\end{abstract}

\pacs{98.80.-k, 95.36.+x, 11.25.Mj}

\maketitle


The experimental discovery that the universe is dominated by a dark energy, which comprises
over $2/3$ of its mass contents, has had profound impact both in cosmology and in the
quest for the microscopic theory of nature. In recent years it has stimulated a remarkable convergence of 
the inflationary paradigm and string theory. The emerging idea of the origins of our universe 
is based on the concept of a `string landscape' \cite{lenny,kklt,douglas},
the myriad of consistent string vacua distinguished by specific values of moduli, which is populated by
the self-reproduction mechanism of eternal inflation. Some of its corners, this framework posits, may yield 
big hospitable universes as our own. 

A particularly important aspect of the landscape approach to describing our universe is how it addresses the cosmological constant problem. The idea is that the cosmological constant varies over the landscape, just like any other low energy Lagrangian parameter of the theory. It can change by nucleation of membranes \cite{brownteitel}, charged under a locally constant $4$-form field whose flux compensates bare vacuum energy \cite{BP,feng}. The membranes are nucleated during inflation, and in some regions of the dynamical landscape may yield a nested system of vacuum bubbles with the vacuum energy inside the bubbles changing across the system of boundaries. Bousso and Polchinski have shown \cite{BP} how a mechanism, generalizing the earlier proposal by Linde \cite{andreipr} and by Brown and Teitelboim \cite{brownteitel}, can be embedded in string theory, in a way which yields a set of states with different charges, but only a tiny difference between their vacuum energy densities, due to incommensurability of the membrane charges sourcing the $4$-forms. They have outlined specific requirements for the corner of the landscape with states where this vacuum energy mismatch is comparable to the residual vacuum energy that would explain current cosmological observations, 
$\Lambda \sim 10^{-12} {\rm eV}^4$. Bousso and Polchinski then showed that the random dynamics of successive membrane emissions during inflation can take the system somewhere in space to a state with so miniscule a vacuum energy, with a large last jump which left enough room for a stage of a slow roll inflation to refill the universe with matter. This then set the stage for the anthropic resolution of the cosmological constant, championed by Weinberg \cite{weinberg}, Linde \cite{lindeq} and Vilenkin \cite{vilenkin}.

This explanation fits the observations, and emerges from the idea of string landscape. Its immediate prediction which could in principle be falsified, is that the dark energy equation of state is $w=-1$. In fact such was the dearth of reasonable explanations of vacuum energy that could be embedded in string theory, that it has been suggested that the dark energy models with local dynamics, such as quintessence, are an unnecessary digression \cite{bousso}. Given the obstacles for accommodating accelerating cosmologies in string theory, both conceptual \cite{witten,banks,hks} and practical \cite{maldan}, it might have indeed seemed that seeking for a serious candidate for quintessence, which really~\footnote{Here, we mean a quintessence model which really comes from string theory, as opposed to various ``stringy motivated" suggestions.} dwells in string theory is in vain.

The aim of this article is to argue this is not so. On the contrary, a mild extension of the arguments employed in the Bousso-Polchinski proposal for relaxing the vacuum energy, by ingredients already present in the landscape framework, yields a candidate low energy theory of quintessence~\footnote{Upon the completion of this work, Fernando Quevedo has informed us of another example of candidate for quintessence which he argued for in several recent talks.}. The key role is played by internal components of higher rank forms. These fields are axion-like at low energies, after compactification. In the presence of internal fluxes in orthogonal subspaces 
they will mix with residual $4$-forms after compactification due to the trilinear Chern-Simons terms, where -- as is usual -- we assume that the dilatonic volume moduli are all stabilized. 
The mixing generates an axion mass term while {\it preserving} the axion 
shift symmetry of the action, which is broken spontaneously once the background solution is chosen 
\cite{giavilen}. When there are more axions, which couple to more $4$-forms in $4D$, the axion mass matrix generated by random fluxes may have ultralight eigenmodes over the landscape, 
if the string scale is low, as invoked by Bousso and
Polchinski in one of the implementations of their mechanism. The reason for the smallness of the mass is, roughly, similar to why there may be a small jump in the absolute value of the cosmological constant between subsequent local vacua, arising from a small mismatch between the charges of different form fields. We illustrate this with explicit examples with few axions and $4$-forms, that can come about if the internal manifold has multiple higher rank forms, as in e.g. type IIB theories. Evaluating the mass eigenvalues, we show that the lightest mass can be comparable to the Hubble scale now, $H_0 \sim 10^{-33} {\rm eV}$. In this case,
the theory has low string scale, $M_s \sim {\rm few} \times 10 \, {\rm TeV}$, and two large dimensions, $L \sim 0.1 \, {\rm mm}$, just like in the simplest large dimensions scenario of \cite{add,aadd}.
The quintessence mass is protected from perturbative corrections in field 
theory by the shift symmetry of the axion effective action. If the ultralight eigenmode does not couple directly to any sector strongly 
coupled at a high scale, the non-perturbative field theory corrections to its potential are also 
suppressed. Moreover, when the compactification length is larger than the string length, the gravitational corrections may remain small too even when the field value approaches $M_{Pl}$, exceeding the effective axion width constant $f_a$ and yielding a very curvy potential. When this happens, the nonperturbative potential is negligible, leaving the residual flat potential generated only by the form fields.
While it is not yet clear that the low energy Standard Model cohabitates with quintessence
in this precise corner of the landscape, it is at least possible to find it in various type IIB compactifications \cite{aadd,antpiol,fernando,cullen}. Since the mechanism which we illustrate is quite generic, it may appear in places where the Standard Model lurks.

Let us now review the dynamics of axions coupled to forms. Start with the simplest case of a single axion mixing with a single $4$-form, via a term $\sim \phi \epsilon^{\mu\nu\lambda\sigma} 
F_{\mu\nu\lambda\sigma}$. The action, which includes minimal coupling to gravity,
consistent with the assumption that all volume moduli are stabilized, is composed of bulk and membrane terms. The bulk term is
\begin{eqnarray}
{\cal S}_{bulk} &=&\int d^4 x \sqrt{g} \left( \frac{M_{Pl}^2}{2} R -\frac12({\nabla}\phi)^2 - 
\frac{1}{48} F_{\mu\nu\lambda\sigma}^2 + \right.\nonumber\\
&+&\left. \frac{\mu}{24}\,\phi\, \frac{\epsilon^{\mu\nu\lambda\sigma}}{\sqrt{g}}
F_{\mu\nu\lambda\sigma} + \ldots \right) \, ,
\label{actionbulk}
\end{eqnarray}
where the ellipsis refer to the matter sector contributions, and $\epsilon^{\mu\nu\lambda\sigma}$ is the Levi-Civita tensor density, as indicated by the explicit factor of metric determinant. The $4$-form field strength 
is the antisymmetric derivative of the $3$-form potential 
$F_{\mu\nu\lambda\sigma} = 4\, \partial_{[\mu} A_{\nu\lambda\sigma]}$. The parameter
$\mu$ has dimension of mass, as required to correctly normalize the bilinear  $\phi \epsilon^{\mu\nu\lambda\sigma}  F_{\mu\nu\lambda\sigma}$. In dimensional reduction $\mu$
is the flux through compact dimensions, as we will see later. The membrane term includes the standard coupling to the $3$-form potential, 
\be
{\cal S}_{brane} \ni \frac{e}{6} \int d^3 \xi \sqrt{\gamma} e^{abc} \partial_a x^\mu  \partial_b x^\nu  \partial_c x^\lambda A_{\mu\nu\lambda} \, ,
\label{actionbrane}
\ee
where the integration is over the membrane woldvolume $\xi^a$ with induced metric $\gamma_{ab}$. We have absorbed numerical factors in the membrane charge $e$, which is normalized to the membrane tension, and may be renormalized by internal volume factors if the membrane is actually a higher-dimensional $p$-brane which wraps some of the compact dimensions, with more details below. The membrane action also includes the membrane kinetic terms, or equivalently, the boundary terms for bulk fields which ensure that the membrane is embedded along woldvolumes which respect canonical bulk boundary conditions. These terms are the Gibbons-Hawking term for gravity, and its analogue for the $4$-form \cite{duff,duncan}. When $\mu=0$ this term is
$\int d^4 x \sqrt{g} ~ \frac16 ~ \nabla_\mu ( F^{\mu\nu\lambda\sigma} A_{\nu\lambda\sigma})$ with our normalizations. However, when $\mu \neq 0$, an extra contribution, 
$- \int d^4 x \sqrt{g} ~  \frac16 ~ \nabla_\mu ( \mu \phi \frac{\epsilon^{\mu\nu\lambda\sigma}}{\sqrt{g}} A_{\nu\lambda\sigma})$ must be added. 
When $\mu$ vanishes, in $4D$ the $4$-form is non-propagating: because
it is completely antisymmetric, its field equations are locally trivial in the bulk, and its value locally constant. In the presence of membranes, however, the $4$-form can change between interior and exterior of the membrane, jumping across its surface. Indeed, setting $\mu=0$ and varying 
(\ref{actionbulk})-(\ref{actionbrane}) with respect to $A_{\mu\nu\lambda}$ yields
\begin{eqnarray}
&&\nabla_\mu F^{\mu\nu\lambda\sigma} = 0 \, ,  ~~~ {\rm in \,  the \, bulk} \, , \\
&&\Delta F_{\mu\nu\lambda\sigma} = e \sqrt{g} \epsilon_{\mu\nu\lambda \sigma} \, , ~~~ {\rm across \, the \,  membrane} \, .
\label{4formeqs}
\end{eqnarray}
Thus the $4$-form is locally indistinguishable from a (positive) contribution to the cosmological constant, which can be reduced by membrane emission in the interior of the membrane. 

There are important differences when $\mu \ne 0$. As already noted in \cite{giavilen}, although still without local propagating modes in this case the $4$-form is {\it not} locally constant. Instead, it is proportional to the scalar field $\phi$, which mixes with it, and so it may vary from place to place. In turn, the scalar field is {\it massive}: the $4$-form background provides an inertia to the scalar's propagation, which by local Lorentz invariance translates into the scalar mass term. Once the background is selected, and the value of
the $4$-form locked to that of $\phi$, the shift symmetry is broken {\it spontaneously}, with the vacuum selection \cite{giavilen}. Still, at the level of the action it remains operative, as can be seen readily from (\ref{actionbulk}):
under $\phi \rightarrow \phi + \phi_0$, the action changes only by a total derivative, and so the local dynamics remains invariant. In fact, although this total derivative could affect the membrane 
action (\ref{actionbrane}), it gets completely cancelled on any physical membrane term by the variation of the boundary term $- \int d^4 x \sqrt{g} ~  \frac16 ~ \nabla_\mu ( \mu \phi \frac{\epsilon^{\mu\nu\lambda\sigma}}{\sqrt{g}} A_{\nu\lambda\sigma})$, possibly leaving only a boundary term at infinity. So actually the theory retains full shift symmetry in the action.

These statements can be simply verified by working explicitly in the action. We can integrate out the $4$-form, because it remains an auxiliary field even when 
$\mu \ne 0$, since it is fully determined by $\phi$ and an integration constant. This integration constant can be recovered by the Lagrange multiplier method \cite{nktop}: first, we recast (\ref{actionbulk}) in the first order formalism, 
enforcing the relation $F_{\mu\nu\lambda\sigma} = 4 \partial_{[\mu} A_{\nu\lambda\sigma]}$ with a Lagrange multiplier. Because of antisymmetry, it takes only one multiplier $q$, and the result is to add the term
\be
{\cal S}_q = \int d^4 x \,   \frac{q}{24} \, \epsilon^{\mu\nu\lambda\sigma} \, \Bigl(
F_{\mu\nu\lambda\sigma} -  4 \partial_{\mu} A_{\nu\lambda\sigma} \Bigr) \, 
\label{lagrang}
\ee
to the action (\ref{actionbulk}). 
Then we can complete the squares in $F_{\mu\nu\lambda\sigma}$ introducing the new variable $\tilde F_{\mu\nu\lambda\sigma} = F_{\mu\nu\lambda\sigma} - \sqrt{g} \epsilon_{\mu\nu\lambda\sigma} (q + \mu \phi)$. The action only depends on $\tilde F_{\mu\nu\lambda\sigma} $ through $\tilde F^2$, 
which therefore yields a Gaussian functional integral and can be dropped as an overall normalization of the partition function. This effectively replaces the $4$-form with its Hodge dual, and enforces the $4$-form equation of motion as a constraint. 
The end result is the effective action describing the $\phi-q$ sector coupled to gravity
\begin{eqnarray}
{\cal S}_{eff} &=&\int d^4 x \sqrt{g} \left( \frac{M_{Pl}^2}{2} R -\frac12({\nabla}\phi)^2 
- \frac12 (q + \mu \phi)^2 + \right.\nonumber\\
&+&\left.\frac16 \frac{\epsilon^{\mu\nu\lambda\sigma}}{\sqrt{g}} \, A_{\nu\lambda\sigma} \, \partial_{\mu} q \right) \, , 
\label{effact}
\end{eqnarray}
where the last term was obtained from an integration by parts, and its total derivative completely cancels against the membrane  terms $ \frac16 ~ \nabla_\mu [ ( F^{\mu\nu\lambda\sigma}
- \frac{\epsilon^{\mu\nu\lambda\sigma}}{\sqrt{g}} \mu \phi ) A_{\nu\lambda\sigma}) ]$ after the shift to the new variable $\tilde F_{\mu\nu\lambda\sigma}$. The charge term (\ref{actionbrane}) then still remains, as it controls the global dynamics of the field $q$. Locally, this field is a constant, as is clear from 
(\ref{actionbulk}) upon variation with respect to the $3$-form $A_{\mu\nu\lambda}$, which yields $\partial q = 0$. In the presence of a membrane, however, the membrane term (\ref{actionbrane}) acts as a source for $\partial q$, and shows that it jumps in the direction along the normal to the membrane,
\be
\Delta q|_{\vec n} = e \, .
\label{qeq}
\ee
This reproduces the boundary condition for the $4$-form (\ref{4formeqs}) in the dual formulation of (\ref{effact}). 

From (\ref{effact}) it is immediately clear that $\phi$ is massive,
with $\mu$ being precisely its mass. The $4$-form in the bulk yields an effective potential
$V = \frac12 (q + \mu \phi)^2$ instead of the pure cosmological constant contribution $\frac12 q^2$. 
In spite of the $\phi$-dependent potential, the shift symmetry $\phi \rightarrow \phi + \phi_0$ is {\it not explicitly broken} in the action. Indeed, the variation of $\phi$ is compensated by the shift of the `field' $q$ according to $q \rightarrow q - \mu \phi_0$, such that both the bulk action (\ref{effact}) and the membrane term remain unchanged! On the other hand, once the vacuum is picked by selecting the solution $q = q_0$,
specified by the membrane sources in the spacetime, the shift symmetry is broken spontaneously,
and the field $\phi$ is massive. In the $\phi$-vacuum, $\phi = -q_0/\mu$, the $4$-form contribution to the vacuum energy is completely cancelled by the scalar field contribution. Hence if the mass $\mu$ is large, greater than the 
Hubble scale of the universe, the field $\phi$ will rapidly roll to the minimum of the potential preventing the $4$-form that it mixes with from participating in the neutralization of the vacuum energy. This could be averted if the axion $\phi$ picks up additional potential terms which stabilize it near $\phi=0$, counteracting the mixing effects and possibly explicitly breaking shift symmetry. Clearly, if this does not occur, only the forms which do not mix with any heavy axions can play a role in the cosmological adjustment of the vacuum energy. In what follows, therefore, we will assume the existence of both forms which do, and which don't mix with axions.

The unbroken shift symmetry in the action (\ref{effact}) implies that a 
{\it massive} field  $\phi$ retains a protective mechanism in perturbation theory which prevents radiative corrections to its mass. Indeed, $\phi$ will couple to other matter only derivatively, and so radiative corrections generated through those couplings will not shift the mass term away from the value induced by the mixing with the $4$-form, as it is the only perturbative term of dimension 2. Further, since the $4$-form remains auxiliary in $4D$ even when $\mu \ne 0$, it does not involve local dynamics that  can change the scalar mass in the framework of $4D$ EFT. Thus as far as perturbation theory is concerned, once $\mu$ is set, it stays put on a fixed background.

That does not imply that the mass $\mu$ is an absolute constant. As we have already noted, in the context of dimensional reduction, which one expects to lead to actions like (\ref{actionbulk}), the parameter $\mu$ is an internal form flux. Let us illustrate this. Consider a simple dimensional reduction of the $4$-form sector in 
$11D$ {\it SUGRA}, on a background which factorizes as a $4D$ spacetime, a three-torus and a four-torus, $M_4 \times T^3 \times T^4$, and take the $3$-form potential $A$ with components $A_{\mu\nu\lambda}(x^\sigma)$ in $M_4$, ${\cal A}_{abc}(x^\sigma)$ on $T^3$, and $\hat A_{ijk}(y^l)$. So, the potential on the three-torus depends only on the spacetime coordinates, whereas the potential on the four-torus is independent of the spacetime location. For the components of $F=dA$, the field equations are
$d^*F = \frac{1}{2} \, F\wedge F$, where the $3$-form potential $A$ is dimensionless, with the $11D$ form action normalized to
$S_{4~form} \sim M_{11}^9 \int ~ {^*} F \wedge F + \ldots$,  with $M_{11}$ the $11D$ Planck mass. Substituting our $3$-form Ansatz first yields
$\nabla_i F^{ijkl} = 0$, which implies that $F_{ijkl} = \mu \epsilon_{ijkl}$ on the four-torus. Then, after straightforward manipulation, defining $\varphi = {\cal A}_{abc}$, and introducing canonically normalized $4D$ fields $\phi = \frac{M_{Pl} \varphi}{V_3 M_{11}^3}$, and $F_{\mu\nu\lambda\sigma} = \frac{M_{Pl}}{\sqrt{2}}
4 \partial_{[\mu} A_{\nu\lambda\sigma]}$, the remaining equations reduce to $\nabla^2 \phi = \mu \, \sqrt{g_{4}} \epsilon_{\mu\nu\lambda\sigma} F^{\mu\nu\lambda\sigma}$ and $\nabla^\mu F_{\mu\nu\lambda\sigma} = \mu \, \sqrt{g_{4}} \epsilon_{\mu\nu\lambda\sigma} \partial^{\mu} \phi$, where the mass parameter $\mu$ is {\it exactly} the internal four-torus magnetic flux of $F_{ijkl}$, up to possibly a combinatorial factor of ${\cal O}(1)$. These are {\it precisely} the variational equations which follow from (\ref{actionbulk}). Hence indeed (\ref{actionbulk}) can be interpreted as a truncation of $11D$ {\it SUGRA}, if all other moduli are stabilized. In fact, there are string theory constructions where such low energy dynamics are known to arise 
\cite{bewitt,shaka}. This also shows that the $4D$ axion mass $\mu$ can change if the magnetic flux in the internal dimensions changes, for example by membrane nucleation. Indeed, if a membrane charged under $F_{ijkl}$ is nucleated, inside the bubble of space enveloped by it the flux, and consequently also the axion mass, will change, to $\mu' = \mu - e$. In other words, the parameter $\mu$ is completely analogous to the variable $q$ which we introduced in the dual formulation of the $4$-form action (\ref{effact}). Just like the vacuum energy, the low energy dynamics of the axion will also be controlled by very different scales in different regions of the Metauniverse, when it is permeated by the many bubbles formed by membrane nucleations \cite{andreiselfrep,donoghue,lenny,BP} (other aspects of eternal inflation were 
discussed in \cite{eternal}). Inflation will ensure that at low energies the universe will in fact be composed of a diverse set of regions with vastly different values of the axion mass.

Clearly, the mass can change in discrete steps. However, an even stronger statement holds: the mass $\mu$ is in fact quantized in the effective $4D$ theory, just like any $4$-form flux. The elegant discussion of this issue is presented by Bousso and Polchinski \cite{BP}. The point is that the classical integration constant which arises in the solution for the $4$-form field strength, $F_{ijkl} = \hat \mu \epsilon_{ijkl}$ can only take discrete values, quantized in the units of the membrane charge. The argument which shows this is similar to the Dirac string construction, and is most readily understood from the viewpoint of the higher-dimensional parent theory, where all the $4$-form field strengths are sourced by membranes or fivebranes. Thus, with our normalization, the quantities $q$ and $\hat \mu$ should be viewed as the integer multiples $q_i$ of the appropriate membrane charges~\footnote{{The generation of a quadratic potential for the axion with discrete values of $q$ is discussed in detail, in the context of 11-dimensional sugra, in~\cite{bewitt}.}},
\be
q_i = n_i \, \frac{e_{11}}{\sqrt{Z_i}}  \, , \label{charges}
\ee
where $Z_i$ are the internal volume factors which depend on the dilatonic  moduli, and $e_{11} = 2\pi M_{11}^3$ is the fundamental membrane charge, normalized to the $11D$ Planck mass $M_{11}$. For the electric 
forms these factors are $Z_e = 2\pi M_{11}^9 V_7 = M_{Pl}^2/2$, while for magnetic forms they are $Z_{m,i}= \frac{2\pi M_{11}^3 V_7}{V_{3,i}^2} = \frac{M_{Pl}^2}{2 M_{11}^6 V_{3,i}^2}$ \cite{BP}. Although these quantization rules were nominally derived in the absence of mixing counterterms which arise from the reduction of the Chern-Simons action, they remain valid in the limit of thin membranes, because for continuos field configurations the integrals of the products of $A$ and $F$ over the thin membrane vanish. 
For the specific application to the case of interest to us, since $\mu$ is the charge of a magnetic $4$-form, these formulas give
\be
\mu = 2\pi \, n \, V_3 M_{11}^3  \, \Bigl(\frac{M_{11}}{M_{Pl}} \Bigr)^2 \, M_{11} \, .
\label{ourcharge}
\ee
The change of $\mu$ when a membrane of unit charge is emitted is $\Delta \mu \sim  V_3 M_{11}^3  \, \Bigl(\frac{M_{11}}{M_{Pl}} \Bigr)^2 \, M_{11}$, and is clearly the smallest when the internal three-torus volume is comparable to the $11D$ Planck scale,
$V_3 M_{11}^3 \sim 1$. The numerical lower bound can be easily estimated by recalling that $M_{11}$ may be as low as the electroweak scale, $M_{11} \ga M_{EW} \sim {\rm TeV}$, which implies that
$\Delta \mu \ga 10^{-16} \, {\rm eV}$. Clearly, in this case the four-torus volume $V_4$ must be large in the units of $M_{11}$ to give the hierarchically large $M_{Pl}$, but to get there one needs linear dimensions
to exceed $M_{11}^{-1}$ by a factor of $\sim 10^{7.5}$.  

By itself, this is not sufficient to make $\phi$ a quintessence field. In the regions of the smallest mass
$\mu_{min} \sim \Delta \mu$, the field $\phi$ would fall out of slow roll in the very early universe, when the temperature is of the order of $T \sim \sqrt{\Delta \mu/H_0} \, {\rm Kelvin} \sim 10^8 \, {\rm Kelvin}$, or around the time of nucleosynthesis. Curiously, this is close to the mass required for a pseudoscalar which could affect supernovae dimming, if it coupled to ordinary electromagnetism, as explained in \cite{ckt}. On the other hand, as Bousso and Polchinski noted, the scale which one gets from a single scalar is also too coarse to provide a plausible mechanism for gradual relaxation of vacuum energy. To address this, they pointed out that parametrically much smaller differences between vacuum energies of different states may be engineered in multi-form frameworks. There, form fields with incommensurate charges give rise to vacua with very different form charges, but tiny variation of the net vacuum energy. Note, however, that the problem with the simple setup above is purely numerical: getting the mass to be as small as the current value of the Hubble scale. Without it, one finds a perfectly reasonable agent for driving cosmic acceleration at a higher scale: an inflaton. We hope to revisit this interesting avenue elsewhere \cite{kspreppy}.

We now argue that multi-field setups also yield small net masses for at least one of the axions. So, imagine that the low energy theory contains {\it several} copies of the axion-form sector in (\ref{actionbulk}), or (\ref{effact}). Such cases may occur in, for example, multi-throat compactifications, where at low energies there is replication of degrees of freedom. Since the throats connect to the bulk of the internal Calabi-Yau manifold, the wavefunctions of fields residing in different throats have an overlap. The kinetic terms for the axions and $4$-forms can be separately rotated to the orthogonal, canonical form, leaving us with mixed coupling terms.
Similar mixing terms will arise from direct compactifications of higher-dimensional theories with more higher-rank forms, such as type IIB string theory. In general, the low energy action found in such constructions will gain the form
\be
S_{couplings} = \int d^4x  \sum_{a,b} \, \mu_{ab} \, \epsilon^{\mu\nu\lambda\sigma} \, 
F^a_{\mu\nu\lambda\sigma} \phi^b \,  \, .
\label{mixedcoupls}
\ee
The matrix $\mu_{ab}$ is the mixing matrix between different forms and axions, and in general it needs not even be square. The low energy axion mass matrix is related to $\mu_{ab}$. It can be obtained quickly by employing the same trick we used to get (\ref{effact}). So, rewrite the action with several axion-form sectors in the first order formalism introducing a Lagrange multiplier for each $4$-form. Then integrate out the $4$-forms. 
The action which remains is 
\begin{eqnarray}
{\cal S}_{eff} &=&\int d^4 x \sqrt{g} \left( \frac{M_{Pl}^2}{2} R -\frac12 \sum_b ({\nabla}\phi^b)^2 +\right. \\
&-&\left. \frac12 \sum_a (q^a + \sum_b \mu_{ab} \phi^b)^2 + \frac{\epsilon^{\mu\nu\lambda\sigma}}{6\,\sqrt{g}} \sum_a A^a_{\nu\lambda\sigma} \, \partial_{\mu} q^a \right) , \nonumber
\label{effactm}
\end{eqnarray}
and so the axion mass matrix is
\be
M_{bc} = \sum_a \mu_{ab} \mu_{ac} \, .
\label{massm}
\ee
If we choose to normalize the matrix 
$\mu_{ab}$ to a selected scale $\mu_0 = \mu/n$ of Eq. (\ref{ourcharge}), on the assumption that this is the smallest such scale in the construction, the matrix $\mu_{ab}/\mu_0$ is a dimensionless matrix. The diagonal entries are given by the combinatorial factors times the internal flux in the units of $\mu_0$, following our discussion leading to (\ref{ourcharge}), whereas the off-diagonal entries measure the coupling of different sectors. For example, in throaty compactifications they are controlled by the ratio of the Calabi-Yau volume $V_{CY}$ to the throat volume $V_{throat}$. Their precise numerical value will depend on the details of the construction, and one expects them to be adjustable parameters depending on where the volume moduli are stabilized. In fact, some of the numerical tunings may be mitigated with more degrees of freedom. Specifically, if the mixing matrix entries arise due to independent internal fluxes, they are multiples of combinatorial factors and possibly large integers. In the phase lattice of such a space, there will be points where some of the eigenvalues are very small, even when the individual matrix elements are much larger than unity, similarly to what occurs in the cosmological constant adjustment of Bousso and Polchinski. 
To be able to say more about such examples, we need to consider a more detailed setup. 

A simple
example is provided by the case with {\it three} axions and {\it three} $4$-forms, but also {\it eight} $3$-forms with internal space fluxes, in type IIB theory. The action for the bosonic sector of type IIB supergravity is, in the Einstein frame, and ignoring the dilaton kinetic terms on the assumption that the dilaton is stabilized,
\begin{eqnarray}
S_{IIB} &=& \frac{1}{2\kappa^2_{10}} \int d^{10} x \sqrt{g}  \left( R - \frac{1}{12} g_s^{-1} H_{3}^2 - \frac{1}{12} g_s 
F_{3}^2 +\right.\nonumber\\
&-&\left. \frac{1}{240} {\tilde F}_{5}^2 \right) + \frac{1}{4\kappa_{10}^2}
\int F_5 \wedge B_2 \wedge F_3
\, ,
\label{iib}
\end{eqnarray}
where $g_s = e^{\phi}$ is the string coupling, $2\kappa_{10}^2 = (2\pi)^7 \alpha'^4$ is given in terms of the string scale $\alpha'$, $H_3=dB_2$ and 
$F_3=dC_2$ the
NS and KR $3$-form field strengths and $2$-form potentials. Similarly, $F_5 = dC_4$ are
the $5$-form field strength and $4D$-form potential, which define the self-dual $5$-form 
$\tilde F_5 = F_5 - \frac12 C_2 \wedge H_3 + \frac12 F_3 \wedge B_2$ where 
${^*}{\tilde F}_5 = {\tilde F}_5$. Assuming that the volume moduli are all stabilized by additional ingredients in the theory, and ignoring their dynamics hereafter, reduce (\ref{iib}) to $4D$ by using the consistent truncation of the form sector by using
\begin{eqnarray}
\begin{array}{ll}
F^1_{\mu\nu\lambda\sigma} = M_{Pl}\, {\tilde F}_{\mu\nu\lambda\sigma\, 5}/{\sqrt{2}}\,, & \phi_1 = M_{Pl} B_{47}/{\sqrt{6 g_s}} \, , \\
F^2_{\mu\nu\lambda\sigma} = M_{Pl}\, {\tilde F}_{\mu\nu\lambda\sigma\, 6}/{\sqrt{2}}\,, & \phi_2 = M_{Pl} B_{48}/{\sqrt{6 g_s}} \, , \\
F^3_{\mu\nu\lambda\sigma} = M_{Pl}\, {\tilde F}_{\mu\nu\lambda\sigma\, 7}/{\sqrt{2}}\,, & \phi_3 = M_{Pl} B_{49}/{\sqrt{6 g_s}} \, , 
\end{array}
\label{formtrunc}
\end{eqnarray}
where all of these fields are taken to depend on the $4D$ coordinates $x^\mu$, and the $3$-forms with internal fluxes only,
\be
F_{ijk} = (2\pi)^2 \alpha' \frac{n_{ijk}}{L_i L_j L_k} \, ,
\label{3fluxes}
\ee
where $(i,j,k)$ take values in the set
\begin{eqnarray}
&&\{(5,6,8), \, (5,6,9),\,  (5,7,8),\, (5,7,9),  \nonumber\\
&&(5,8,9),\,(6,7,8),\, (6,7,9),\, (6,8,9)\} \, , \label{set}
\end{eqnarray}
$n_{ijk}$ are the units of flux in $F_{ijk}$ in the directions parameterized by $(i,j,k)$ and $L_i$ are the 
sizes of the dimensions supporting the $F_{ijk}$ flux. One can check directly that (\ref{3fluxes}) obey the form field equations following from (\ref{iib}), and so this truncation is consistent, if the volume moduli are stabilized. Clearly, there are other possibilities from truncations similar to the one displayed here, starting with a trivial exchanging of dimensions used here for those which were ignored. We leave the general case aside, to be addressed in future work, since this one example is sufficient for illustrative purpose. The dimensionally reduced $4D$ effective Lagrangian becomes
\ba
{\cal S}_{eff} &=&\int d^4 x \sqrt{g} \left.( \frac{M_{Pl}^2}{2} R -\frac12 \sum_{b=1}^3 ({\nabla}\phi^b)^2 +\right. \\
&-& \left.\frac{1}{48} \sum_{a=1}^3 \bigl(F^a_{\mu\nu\lambda\sigma}\bigr)^2 + \frac{1}{24} \frac{\epsilon^{\mu\nu\lambda\sigma}}{\sqrt{g}}  \sum_{a,b=1}^3 
\mu_{ab} F^a_{\mu\nu\lambda\sigma} \phi^b  \right) \, .\nonumber
\label{effactmn}
\ea
Let us now choose four of the compact dimensions to be string scale, $L_4= \ldots = L_7 = \sqrt{2\pi \alpha'}$, and two to be larger than the string scale: $L_8 = L_9 = L \gg \sqrt{2\pi \alpha'}$. Further, let us pick the
fluxes such that $n_{589} = -n_{689} =-1$, $n_{579} = n_{568} = n-1$,
$n_{569} = n_{578} = n_{679} = -n$ and $n_{678} = n+1$. Then, the mixing matrix $\mu$ becomes
\be
\Bigl( \mu_{ab} \Bigr) = 
\pmatrix{
\varepsilon^2 & n\varepsilon & (n+1)\varepsilon \cr
\varepsilon^2 & (n-1)\varepsilon & n\varepsilon  \cr
0 & n\varepsilon & (n-1)\varepsilon }  \mu_0 \, ,
\label{mixm}
\ee
where $\mu_0 =  \sqrt{\frac{{3\pi g_s}}{2 \alpha'}}$ and $\varepsilon = \sqrt{2\pi \alpha'}/{L}$. The eigenvalues of the mass matrix $M = \mu^T \mu$ are the roots $\lambda = m^2/\mu_0^2$ 
of the cubic $P_3(\lambda) = \lambda^3 - 6n^2 \varepsilon^2 \lambda^2 + 8n^2 \varepsilon^4 \lambda - \varepsilon^8$ (where we have kept only the leading terms in the  limit $\varepsilon\ll 1$, $n\gg 1$).
Since the matrix $M$ is real-symmetric, the characteristic polynomial must have three real roots. To find the smallest one, one could in principle use the Cardano formulas for the roots of a cubic \cite{wiki}. However, a quicker method is to inspect the graph of $P_3$ and realize that due to the signs of the four terms, the smallest root is {\it i)} positive and {\it ii)} controlled by the cancellation between the linear term and the constant. In the limit $n \gg 1$ and $\varepsilon \ll 1$ which we are interested in, the smallest of the three roots is
\be
m^2_{min} 
\simeq \frac{\varepsilon^4}{8 n^2} \mu_0^2 \simeq \frac{3\pi^3 g_s \alpha'}{4 L^4} \frac{1}{n^2}  \, .
\label{mmin}
\ee
Since it is positive, there are no tachyons in the spectrum. In fact, this should have been expected all along, since we know that we can rewrite the action (\ref{effactmn}) in the form (\ref{effactm}), where the potential is a sum of squares, implying that none of the mass eigenmodes are tachyonic. As a matter of fact, the other two roots, by a similar reasoning yielding (\ref{mmin}), will obey $m^2 \ga \varepsilon^2 \mu_0^2$ and
$m^2 \ga n^2 \varepsilon^2 \mu^2_0$, being determined by a different interplay of the terms in the cubic. 
Thus, they may also end up being parametrically smaller than $\mu_0$. We can now use both 
small $\varepsilon$ {\it and} small $1/n$ to render $m_{min}$ much lighter than $\mu_0$. However, we can't make $n$ as large as we wish, since the internal fluxes after dimensional reduction contribute to the effective $4D$ cosmological constant, as $\Lambda_4 \ni g_s F_3^2$. In fact, if we take the correct UV cutoff of the effective IIB {\it SUGRA} to be set by the string scale, above which we need full string dynamics, we should require that $g_s F_3^2 \la \frac{1}{2\pi \alpha'}$. Evaluating the flux for our truncation and substituting in this formula, we find
\be
n^2 \la \frac{1}{24\pi^2 \varepsilon^2  g_s} \simeq  \frac{L^2}{48\pi^3 \alpha' g_s} \, .
\label{fluxbound}
\ee
Using the Gauss law relating $4D$ Planck mass, the string scale and the compactification volume, 
$M_{Pl}^2 = V_6/\kappa_{10}^2$, yields $L^2 = 16\pi^5 M_{Pl}^2 \alpha'^2$, and therefore
\be
m^2_{min} \simeq \frac{3 g_s}{2^{10}\pi^7 M_{Pl}^4 \alpha'^3} \frac{1}{n^2} \ga \frac{9g_s^2}{2^{10} \pi^9 M_{Pl}^6 \alpha'^4}  \, .
\label{mminf}
\ee
Hence for consistency, $m_{min}$ cannot be dialed down below the lower bound in Eq. (\ref{mminf}).
On the other hand, this mode will be quintessence as long as $m_{min} < H_0$. For this to be possible, we need to ensure that the string coupling is smaller than the critical value 
\be
g_{s~*} = \frac{32}{3} \pi^{9/2} M_{Pl}^3 \alpha'^2 H_0 \simeq 10^{-2} \, \Bigl(\frac{{\rm eV}}{M_s} \Bigr)^2 \, \Bigl(\frac{M_{Pl}}{M_s} \Bigr)^2  \, ,
\label{mminfq}
\ee
where $M_s$ is the string scale. Now, to ensure that we get reasonable $4D$ phenomenology, 
with $M_{Pl} \sim 10^{18} \, {\rm GeV}$, and $4D$ gravity valid down to millimeter distances, we need to take $L \la 0.1 \, {\rm mm}$, which implies $M_s \ga {\rm few} \times 10 \, {\rm TeV}$. This also guarantees that it is easy to stay in the regime where type IIB is perturbative, since then $g_{s~*} \la 1$ and so we can take any $g_s < 1$. In fact, we can take $g_s \sim 10^{-3}$ which will guarantee that the higher-dimensional Planck mass is somewhat greater than the string scale, $M_{10} \sim M_s/g_s$. For these parameters, somewhere in the landscape of the theory spanned by the volume moduli and internal fluxes,
$n$ will fall in the right regime for the lightest axion mass to be $\la H_0$, so that it could remain in slow roll until throughout the cosmic history to date. 

So far, we have neglected the issue of nonperturbative corrections to the low energy action from gauge and gravitational sectors. In fact, although the shift symmetry provides protection to the axion from perturbative corrections arising from matter that axion couples to, it is explicitly broken by nonperturbative effects. These yield instanton-induced effective potentials, $V_{eff} \sim \sum_n \lambda_n^4 \cos(2n\phi/f_\phi)$, where $f_\phi$ is the axion decay constant, and $\lambda_n$ are dynamically generated scales in the instanton expansion, typically related to the UV cutoff via $\lambda_1 \sim M e^{-\alpha/g}$ and 
with $\lambda_{n>1} < \lambda_1$ (see, e.g. \cite{tometc,nflat}). In QCD, $\lambda_1$ happens to be the QCD scale $\Lambda_{QCD}$, but there are examples where it can be vastly different from a characteristic scale of the low energy theory whose gauge sector yields the potential. Now, in string theory it is very difficult to obtain large axion decay constants obeying 
$f_\phi \ga M_{Pl}$. On the other hand, in much of the quintessence model building, such scales are necessary, since {\it i)} the axion {\it vev} needs to be $\ga M_{Pl}$ in order to yield at least an efold of late acceleration, and {\it ii)} the potential must remain flat enough for this to occur, so that the higher order terms in the Fourier series for $V_{eff}$ remain negligible for all 
$\phi \ga M_{Pl}$. In the case we have described (and in contrast to the more usual models of axion quintessence, such as those discussed e.g. in~\cite{svrcek}), the instanton terms aren't needed to get the axion mass! Indeed, even if $f_\phi < M_{Pl}$, and the higher order terms aren't negligible, as long as 
$\lambda_n^4 < m_{min}^2 f_{\phi}^2$ the instanton mass term is small compared to the mass term induced by the mixing with $4$-forms. So to have a working candidate quintessence, one needs not only to select the right region in the landscape, but also to carefully pick the couplings of the lightest axion to the matter sector. Yet, this is, at least in principle, a problem which is often encountered in the landscape model building, and presumably can be addressed. 

Similar concerns arise when one encounters gravitational effects~\footnote{Perturbative graviton loop effects remain fully under control as they can only yield terms $\sim m^2_{min} R$ and $\sim {V_{eff}^2}/{M_{Pl}^4}$ which remain tiny even for large values of the scalar field {\it vev} \cite{andreirad}.}. These effects also yield effective potentials given by harmonic series, but with coefficients proportional to the exponential of instanton action. When we compactify the theory with dimensions which are larger than the fundamental scale, the actions will rapidly grow in the units of the string length. In fact, taking the internal dimensions to only exceed the fundamental scale by one order of magnitude will yield actions of the order of $S \sim 10^d$ where $d$ is the number of compact dimensions. Ensuring this is ${\cal O}(1000)$ or more will render the relevant normalization factors small enough to be ignored in the reckoning with dark energy.

Our discussion so far has centered on the existence of an ultralight axion quintessence. Once it's there, how does it actually come to be dark energy at late times? As in the Bousso-Polchinski scenario, most of the bare vacuum energy in our part of the inflating metaverse should be cancelled by the $4$-forms which do not mix with the axions. For this purpose, one needs to have a number of such forms in order to ensure that the bare vacuum energy in some states can be cancelled with the precision set by the value of the allowed vacuum energy now, $10^{-12} \, {\rm eV}^4$. When the string scale is very low, this can be accomplished with ${\cal O}(10)$ form fields \cite{BP}. In the course of cosmic evolution of our universe, the membranes are emitted during inflation, eventually reducing the net vacuum energy inside the sequence of inflating bubbles down to the presently acceptable value. Part of the effective vacuum energy may also come from the fluxes of the $4$-forms which do mix with the axions. Further, in the very least the light axions will thermally drift around over their domain of definition, and will certainly not rest in the low energy vacua. In fact, the low energy vacua may not even be defined yet, as their actual location is set by the background $4$-form fluxes, which may yet change by membrane emission, as is clear from Eqs. (\ref{effactm}), (\ref{effactmn}). Indeed, the potential for the axion multiplet is $V_{eff} = \frac12 \sum_{a} (q^a + \sum_b \mu_{ab} \phi^b)^2$. After diagonalization, the lightest direction has the effective potential $V_{lightest} = \frac 12 m_{min}^2 (\phi + q_{eff}/ m_{min})^2$, where $q_{eff}$ is a linear combination of the $4$-form fluxes which mix with the axions. Both $q_{eff}$ and $\phi$ may scan their full range of allowed values~\footnote{Examples of fields $\phi$ defined over a large domain, $> M_{Pl}$, in string theory were provided recently in \cite{eva}.}. Generally, 
$q^2_{eff} \gg M_{Pl}^2 H_0^2$, and so a part of $V_{eff}$ will still be cancelled by the forms which do not mix with the axions. It is then sufficient that in some inflating bubbles the final state vacuum energy, involving this linear combination and the additional, unmixed $4$-forms, acquires 
$\phi + q_{eff}/ m_{min} \ga M_{Pl}$. The residual vacuum energy can be $M_{Pl}^2 H_0^2$ and the field will sit in slow roll to the present time, suspended on the shallow potential set by $m_{min}$, with the right value to become the dominant component of dark energy now, and provide an efold or so of accelerated expansion as required by observations. Note, that as the field eventually rolls to its minimum $\phi = - q_{eff}/m_{min}$ and compensates the $4$-form contribution to the vacuum energy, the
leftover vacuum energy might be negative. This means that the universe could collapse in a distant future, realizing a scenario discussed in \cite{lindeq,lindelin,glv}.

In sum, in this work we have argued that the string landscapes may naturally accommodate degrees of freedom which can play role of quintessence. These modes are components of higher-dimensional forms, which mix with $4$-forms in $4D$  theory after compactification. For low string scale, and large extra dimensions, there may be sufficiently light axions which can be quintessence now, with masses
$m \la H_0 \sim 10^{-33} \, {\rm eV}$, that come about thanks to incomplete cancellations between large fluxes of forms much like in the mechanism for canceling vacuum energy of \cite{BP}. The axion shift symmetry protects the quadratic potential against quantum contributions, guaranteeing the flatness of the potential and the absence of an $\eta$ problem. We have provided an explicit example using type IIB theory on a space with two large dimensions and string scale $\sim {\rm few} \times 10 \, {\rm TeV}$, where we assumed the volume moduli to be stabilized. While so far such compactifications haven't seemed exactly realistic from the point of view of low energy particle physics, there is an effort underway searching for type IIB compactifications with only two large dimensions \cite{fern}. Moreover, the main ingredients of the mechanism are sufficiently generic that they may arise in other setups too. It would be interesting to search for the corners of the landscape where the Standard Model may coexist with quintessence modes. 
Further, it is also interesting to classify more precisely cosmological signatures of the quintessence dynamics, as it may accommodate discretely variable mass due to membrane emission. We hope to return to these issues elsewhere.

\vskip.5cm

{\bf \noindent Acknowledgements}

\smallskip

We would like to thank Savas Dimopoulos, Shamit Kachru, Fernando Quevedo and Alessandro Tomasiello for 
valuable discussions. LS thanks the UC Davis HEFTI program for hospitality during the inception of this work. The work of NK is supported in part by the DOE Grant DE-FG03-91ER40674. 
The work of LS is partially supported by the U.S. National Science Foundation grant PHY-0555304.


\end{document}